\documentclass{aa}  

\def\cxo {\emph{Chandra}}
\def\swift {\emph{Swift}}

\def\src {RX\,J0806.3+1527}
\def\flux {\mbox{erg cm$^{-2}$ s$^{-1}$}}
\def\lum {\mbox{erg s$^{-1}$}}
\def\nh {$N_{\rm H}$}
\usepackage{graphicx,graphicx,times,multirow}
\usepackage[]{color}
\usepackage{txfonts}
\begin{document} 

   \title{\swift\ X-ray and ultraviolet observations of the shortest orbital period double-degenerate system \src\ (HM\,Cnc)}
\titlerunning{\swift\ X-ray and ultraviolet observations of  \src}

\author{Paolo~Esposito\inst{1}
\and Gian~Luca~Israel\inst{2}
\and Simone~Dall'Osso\inst{3}
\and Stefano~Covino\inst{4}}

\institute{Istituto di Astrofisica Spaziale e Fisica Cosmica - Milano, INAF, via E. Bassini 15, I-20133 Milano, Italy\\
\email{paoloesp@iasf-milano.inaf.it}
\and Osservatorio Astronomico di Roma, INAF, via di Frascati 33, I-00040 Monteporzio Catone (Roma), Italy
\and Theoretische Astrophysik, IAAT, Eberhard Karls Universit\"at T\"ubingen, Sand 1, D-72076 T\"ubingen, Germany
\and Osservatorio Astronomico di Brera, INAF, via E.~Bianchi 46, I-23807 Merate (LC), Italy}

   \date{Received 20 September 2013 / Accepted 26 November 2013}

 \abstract{The system \src\ (HM\,Cnc) is a pulsating X-ray source with 100 percent modulation on a period of 321.5 s (5.4 min). This period reflects the orbital motion of a close binary  consisting of two interacting white dwarfs. Here we present a series of simultaneous X-ray (0.2--10 keV) and near-ultraviolet (2600 \AA\ and 1928 \AA) observations that were carried out with the \swift\ satellite. In the near-ultraviolet, the counterpart of \src\ was detected at flux densities  consistent with a blackbody with temperature  $(27\pm8)\times10^3$ K.
We found that the emission at 2600 \AA\ is modulated at the 321.5-s period with the peak ahead of the X-ray one by $0.28\pm0.02$ cycles and is coincident within $\pm0.05$ cycles with the optical. This phase-shift measurement confirms that the X-ray hot spot (located on the primary white dwarf) is at about $80\degr$--$100\degr$ from the direction that connects the two white dwarfs. Albeit at lower significance, the 321.5-s signature is also present in the 1928-\AA\ data; at this wavelength, however, the pulse peak is better aligned with that observed at X-rays. We use the constraints on the source luminosity and the geometry of the emitting regions to discuss the merits and limits of the main models for \src.
}

   \maketitle
%

\section{Introduction}

The X-ray emission of \src\ (also known as HM\,Cnc) is modulated on a period of 321.5 s (5.4 min) with virtually no emission for about half of the cycle \citep{ipc99}. The optical counterpart of \src\ has been identified as a blue ($V\sim21.1$ mag / $B\sim20.7$ mag) object with a $\sim$15 percent (pulsed fraction) pulsation at the same period, which leads the X-ray pulsation by $\Delta\phi\sim0.2$ cycles \citep{israel02,ramsay02,barros07}. X-ray and optical monitoring of \src\ have shown that its period is decreasing at a rate of about 1\,ms per year (about $-4\times10^{-11}$ s s$^{-1}$; see \citealt{icd04,strohmayer03,strohmayer05,hakala04,barros07}). No stable periodicities other than the 321.5-s one (and its harmonics) have been observed in this source.

The properties of \src\ and its `twin' RX\,J1914.4+2456 (V407\,Vul), which displays a 569-s (9.5-min) modulation and overall similar X-ray and optical properties, are interpreted in the framework of a double-degenerate binary that hosts two interacting white dwarfs (WDs), where the observed pulsation is due to orbital motion (see \citealt{cropper04,solheim10} for reviews). \citet{roelofs10} showed in \src\ a clear modulation of He emission lines in both radial velocity and amplitude on the 5.4-min period (see also \citealt{mason10}). This confirmed that \src\ is a double-degenerate system and the binary with the shortest known orbital period. As such, it is expected to be a strong source of gravitational waves (GWs), which will be detectable with GW space observatories (e.g. \citealt{nelemans13}).

Several models have been proposed for \src\ and RX\,J1914.4+2456. Among these, mass transfer from a Roche-lobe-filling WD to either a magnetic (polar-like) or a non-magnetic (Algol-like) accretor have been proposed. In the latter model, also known as a `direct impact' accretion model \citep{nelemans01,marsh02}, which assumes a light companion ($\sim$0.14 $M_\odot$), a disk would not form since the minimum distance from the centre of the donor is smaller than the size of the accretor, resulting in the stream directly hitting the surface of the accreting WD. The former scenario \citep{cropper98} is similar to that of polars, where the magnetic field of the accreting WD inhibits the formation of a disk and matter reaches the magnetic polar cap. Models invoking accretion predict, quite generically, orbital widening for two degenerate WDs, in contrast to what is observed in \src\ and RX\,J1914.4+2456. Possible solutions to this difficulty have been proposed \citep{deloye06,dantona06,kaplan12} but also radically different scenarios were elaborated to solve the issue.

The main alternative to accretion models involves a magnetic primary WD and a (non-magnetic) secondary that does not fill its Roche lobe; if the spin period of the primary is not synchronous with the orbital period, then the secondary crosses the primary's magnetic field as it moves along the orbit, and the resulting electromotive force drives an electric current between the two WDs (assuming the presence of ionised material between them), whose dissipation heats the polar caps on the primary. This model, analogous to that proposed for the Jupiter--Io system, is known as the `unipolar inductor' (UI) or 'electric star' model \citep{wu02,wu09,dallosso06,dallosso07,israel09}. Recently, \citet{lai12} highlighted a critical inefficiency of this mechanism in dissipating enough energy to match the X-ray emission of the ultracompact double-WD binaries. Adopting different (but realistic) values for the relevant parameters, this can be partially relieved, although still the model might need some significant improvement beyond the current simple picture (this is discussed in more detail in Sect.\,\ref{disc}).

Here we report on the results obtained from the analysis of a series of observations of \src\ carried out with the X-Ray Telescope (XRT) and the Ultra-Violet/Optical Telescope (UVOT) aboard the \swift\ satellite. These observations allowed us to detect \src\ in the near-ultraviolet (UV) and to perform a timing study simultaneously at X-ray and UV wavelengths. 

\section{Observations and data analysis}

The \swift/XRT \citep{burrows05} uses a front-illuminated CCD detector sensitive to photons between 0.2 and 10 keV. Two main readout modes are available: photon counting (PC) and windowed timing (WT). The PC mode provides two dimensional imaging information and a 2.507-s time resolution. In WT mode only one-dimensional imaging is preserved, achieving a time resolution of 1.766 ms. 
The \swift/UVOT \citep{roming05} is a 30-cm modified Ritchey-Chr\'etien reflector coupled to a microchannel-plate intensified CCD detector; the set of filters covers the range 1700--6000 \AA. The UVOT operates in two data-taking modes. In image mode, the counts are accumulated into an image and only the start and stop times of the exposure are recorded. In event mode, the temporal and positional information of each photon are saved. In this way, the photons can still be stacked into images, but the data can also be used to investigate variability at short time-scales (the timing resolution is equal to the readout time, 11.0322 ms for the full CCD).

From 2005 September to 2008 September, \swift\ observed \src\ nine times. Since most observations were carried out  with the 2600 \AA\ UVOT  $uvw1$ filter (33 ks) to further extend the study of \src\ to short wavelengths, we performed additional observations with the 1928 \AA\ $uvw2$ filter  in March/April 2010 for a comparable exposure time (26~ks). A summary of the datasets used for this work is given in Table~\ref{log}. The XRT and UVOT data were processed and filtered with standard procedures and quality cuts\footnote{See \url{http://swift.gsfc.nasa.gov/docs/swift/analysis/} for more details.} using \texttt{ftools} tasks in the \texttt{heasoft} software package (v.~6.11) and the calibration files in the 2012-02-06 \texttt{caldb} release.  

\begin{table}
\center
\caption{Journal of the \swift\ observations of \src. The XRT data in WT mode, as indicated by the square brackets, were not used for spectral or timing analysis. The star indicates the UVOT exposures in image-only data-taking mode.}
\label{log}
\begin{tabular}{@{}llcc}
\hline
\hline
Obs. ID & Date & XRT & UVOT \\
 & & exp. (ks) & exp. (ks) / filter \\
\hline
35180001& 2005 Sep 10--12 & 13.9\phantom{1} & 14.1 / $uvw1$\\
35180002& 2005 Sep 23 & 2.0 & \phantom{1}2.1 / $uvw1$\\
35180003& 2005 Nov 30 & 2.0  &\phantom{$4.1$} --\phantom{$uvw$} \\
35180004& 2005 Nov 30 & 13.4\phantom{1} & 13.7 / $uvw1$\\
35705001& 2006 Sep 11 & 1.4 & 1.0 / $b$, $v$$^*$\\
35705002& 2007 Mar 18 & 4.3 & 4.3 / $b$, $v$$^*$\\
37762001& 2008 May 31 & 1.2 & \phantom{1$^*$}1.2 / $uvw1$$^*$\\
35180005& 2008 Jun 02 & 1.7 & \phantom{1}1.7 / $uvw1$\\
37762002& 2008 Sep 16 & 0.3 & \phantom{1$^*$}0.3 / $uvw1$$^*$ \\
37762003& 2010 Mar 30 & [7.9]& \phantom{1}7.9 / $uvw2$ \\
37762004& 2010 Apr 02 & 9.5 & \phantom{1}9.5 / $uvw2$ \\
37762005& 2010 Apr 07 & 8.6 & \phantom{1}8.6 / $uvw2$ \\
\hline
\end{tabular}
\end{table}

\subsection{X-ray telescope spectroscopy}\label{xrtspec}
\src\ is clearly detected up to $\sim$1 keV in every XRT exposure but we only considered PC data,  given the relatively low count rate (about 0.025 counts s$^{-1}$), which have an intrinsically higher signal-to-noise ratio with respect to the WT ones. We selected the source events within a 20-pixels radius (1 XRT pixel $\simeq2.36$ arcsec), while the background was estimated from an annular region with radii 50 and 80 pixels. For the spectral analysis, we combined the datasets since all count rates were consistent ($\chi^2$ probability of constancy: 98.9\%),  and extracted a cumulative spectrum. This resulted in a total exposure of $\sim$57.9 ks and about $1440\pm40$ net counts in the 0.3--1 keV range.
The data were grouped so as to have a minimum of 20 counts per energy bin and fed into the \texttt{xspec} fitting package (v.~12.7). The ancillary response file was generated with \texttt{xrtmkarf}, and it accounts for different extraction regions, vignetting and point-spread function corrections. 

Previous  studies indicate that the X-ray flux of \src\ is moderately variable on long time-scales ($\approx$30\%), apart from the 5.4-min modulation, and that its spectrum is consistent with an absorbed blackbody (BB; e.g. \citealt{icd04,strohmayer05,strohmayer08}), so we fit this model to the data [$\chi^2_\nu = 1.29$ for 29 degrees of freedom (dof)]. The absorption column is $N_{\mathrm{H}} = (1.8\pm1.3)\times 10^{20}$ cm$^{-2}$, and the inferred temperature corresponds to $kT = (64\pm3)$~eV (here and in the following all uncertainties are at 1$\sigma$ confidence level). The 0.3--1 keV average observed flux is $1.10^{+0.02}_{-0.03}\times10^{-12}$ \flux. The corresponding unabsorbed flux is  $1.5^{+0.2}_{-0.3}\times10^{-12}$ \flux.

\subsection{Ultraviolet/optical telescope photometry}

The analysis was performed (for each filter) on the stacked images from the whole campaign with the \texttt{uvotsource} task, which calculates the magnitude through aperture photometry within a circular region and applies specific corrections due to the detector characteristics. All magnitudes are in the natural UVOT photometric system (Vega, see \citealt{poole08} for more details and \citealt{breeveld10,breeveld11} for the most updated zero-points and count-rate-to-flux conversion factors).  Since \src\ is a relatively faint source in the UVOT range, we adopted an extraction radius of 3-arcsec and applied the corresponding aperture corrections.\footnote{See\\ \texttt{http://heasarc.nasa.gov/ftools/caldb/help/uvotsource.html}.}

The UVOT exposures with the optical filter $v$ (central wavelength 5468 \AA, FWHM 769 \AA; see Table~\ref{log}) yielded no detection of \src\ with a limit $v>20.7$ mag (total exposure: 2.7 ks). In the $b$ filter (central wavelength 4392 \AA, FWHM 975 \AA), the source is marginally detected (3.5$\sigma$ significance) at $b=(21.0\pm0.3)$ mag (total exposure: 2.5 ks). Both values are consistent with past measurements from various ground-based telescopes \citep{israel02,ramsay02}. 

Most of the UVOT exposures were, however, taken with the near-UV $uvw1$ (32.7 ks) and $uvw2$ (25.6 ks) filters, where the source was detected at higher than $70\sigma$ and $100\sigma$ confidence levels, respectively. The observed apparent magnitude in the $uvw1$ band (central wavelength 2600 \AA, FWHM 693 \AA) is  $uvw1=(18.50\pm0.08)$ mag, including a systematic error of $\pm$0.03 mag for uncertainties in the photometric zero point and flux conversion factor. This translates into a flux density of $(1.6\pm0.1)\times10^{-16}$ \flux\ \AA$^{-1}$. We used the relation by \citet{predehl95} to derive the optical extinction from the measured \nh\ and we find $A_V=0.10\pm0.07$ mag. We note that this value is consistent with the total line-of-sight optical extinction $A_V=0.094$ mag that is estimated from background infrared (IR) emission \citep{schlegel98}. Using the extinction curves by \citet{fitzpatrick07}, we obtained a  de-reddened flux density $F_{uvw1_{\circ}}=1.9^{+0.5}_{-0.3}\times10^{-16}$ \flux\ \AA$^{-1}$. We followed the same procedure for the $uvw2$ data (central wavelength 1928 \AA, FWHM 657 \AA). The  observed magnitude is $uvw2=18.02\pm0.09$ mag (including a $\pm$0.03 mag systematic uncertainty), which corresponds to a flux density of $(3.3\pm0.3)\times10^{-16}$  \flux\ \AA$^{-1}$. After correcting for the appropriate extinction, the de-reddened density of flux is $F_{uvw2_{\circ}}=4.3^{+1.3}_{-1.0}\times10^{-16}$ \flux\ \AA$^{-1}$. 

\subsection{Timing analysis}\label{timing}

\begin{figure}
\center
\resizebox{\hsize}{!}{\includegraphics[angle=-90]{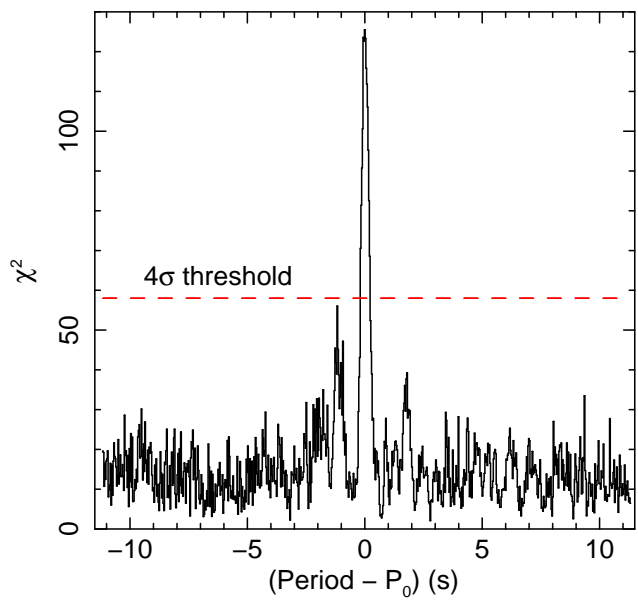}}
\caption{\label{rp_uvw1}
Rayleigh periodogram for the UVOT/$uvw1$ light curves; the offset $P_0$ is the period measured in the simultaneous XRT data. The 321.5-s modulation is clearly detected (see text for details). The 4$\sigma$ threshold is also reported (stepped line).}
\end{figure} 

\indent For the timing analysis, the photon arrival times were corrected to the Solar system barycentre with the \texttt{barycorr} task and using the \cxo\ source position ($\rm RA =08^h06^m22\fs92$, $\rm Dec. =+15\degr27'30\farcs9$, epoch J2000; $0\farcs7$ accurate), as provided by \citet{israel03}. For UVOT, we could only use the time-tagged event-mode data, and for XRT, we used only the PC data (see Table~\ref{log}).

With an X-ray pulsed fraction of $\sim$100 percent, the signal at $\sim$321.5 s is easily detected in every XRT dataset with a long-enough exposure. By means of these new X-ray datasets, we were able to refine our previous coherent timing solution \citep{icd04,idm05,israel09}. According to our analysis, the phase-coherent best inferred X-ray period and period derivative are $P = 321.5303822(13)$ s and $\dot{P} = -3.6718(13)\times10^{-11}$ s s$^{-1}$; MJD = 52619.0000 (TDB time) was used as reference epoch for the period derivative.
\begin{figure}
\center
\resizebox{\hsize}{!}{\includegraphics[angle=-90]{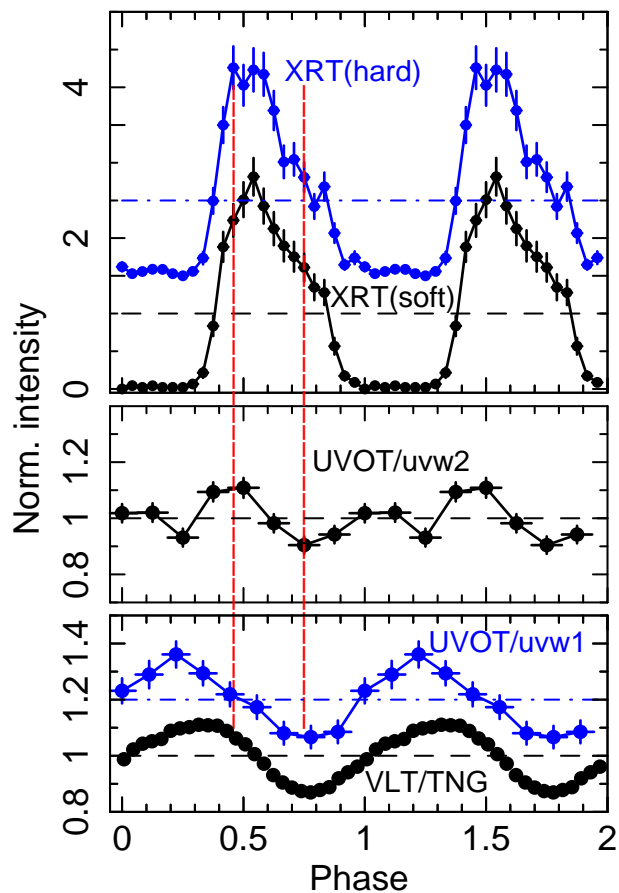}}
\caption{\label{efold} Pulse profiles of \src\ (from the top to the bottom) in the X-rays (upper panel; XRT for $E>0.4$ keV, blue, and $E<0.4$ keV),  UVOT/$uvw2$ (middle panel), UVOT/$uvw1$ (blue) and optical VLT/TNG (lower panel) data. For each profile, the dashed or dot-dashed line indicates the normalized average intensity. Note that the hard XRT and UVOT/$uvw1$ profiles are shifted for display purposes.}
\end{figure}

We performed a search for coherent modulations around the X-ray period in the  UVOT/$uvw1$ data by means of a Rayleigh periodogram. Figure\,\ref{rp_uvw1} shows the result of the search: a significant peak, about 9$\sigma$ for 10$^{3}$ trial periods, is clearly detected in the periodogram, which testifies to the presence of a strong modulation at a period of $321.44\pm0.15$ s (1$\sigma$ c.l.). This is  consistent with that detected in the simultaneous X-ray dataset. A similar result was obtained for the  UVOT/$uvw2$ dataset, though the peak significance is much lower than in the $uvw1$ filter, as it ranges from about 4$\sigma$ to 5.5$\sigma$. The former value was found by performing a search over 10$^{3}$ trial periods (covering approximately a $\pm$10-s period interval around the X-ray value) and the latter focused the search in a narrow period interval around the UVOT/$uvw1$ peak and assumed the $\dot{P}$ value from the updated timing solution (one independent trial). The resulting background-subtracted X-ray and UV \swift\ total light curves folded on the update phase-coherent solution are shown in Fig.\,\ref{efold} with the VLT and Telescopio Nazionale Galileo (TNG, La Palma) optical one \citep{idm05}. The reduction of the optical data have been carried out following the recipe outlined in \citet{israel02,icd04,idm05}.

The pulsed fraction, defined as the semi-amplitude of the sinusoidal modulation divided by the mean source count rate, is $(13\pm2)$ percent in the $uvw1$ band. In contrast to the X-rays (where there is no detected emission for roughly half of the period and the pulse shape has a marked `saw-tooth profile'), the optical profile is almost sinusoidal. The $uvw1$ pulse shape is intermediate: While less dramatic than the X-ray one, it is not purely sinusoidal. The addition of a second harmonic gives a Fisher-test chance probability of $\sim$$4\times10^{-3}$, which denotes that the additional harmonic is significant at the 2.9$\sigma$ confidence level.

A substantial phase shift between the XRT and $uvw1$ optical profiles is also evident in our results. To accurately determine the value of this shift, the $uvw1$ and X-ray pulse profiles were cross correlated, and the resulting peak in the cross-correlation function was fitted with a Gaussian. This yielded a phase lag of $\Delta\phi=0.28\pm0.02$ with the X-ray peak following the $uvw1$ one. When considering only the separation between the maxima of the $uvw1$ and X-ray folded profile, we obtain a value of the phase lag $\Delta\phi=0.20\pm0.03$ (note, however, that this measurement is more dependent on the binning). We stress that no issues related to absolute phase alignment affect the results of this analysis, since all times are assigned with the same spacecraft clock. We repeated the analysis for the VLT/TNG optical data and found that the optical and $uvw1$ pulsations are virtually coincident ($\Delta\phi<0.05$).

The 321.5-s modulation appears to be fainter in the $uvw2$ dataset with a pulsed fraction of about $(7.4\pm1.2)$ percent and a double-peak shape. Interestingly, the highest peak appears almost in phase with the X-ray emission at a variance with both the optical data and the UVOT data in the $uvw1$ filter. On the other hand, the main minimum is approximately coincident with the minima in the $uvw1$ and optical bands, thus hinting at a fundamental connection between them. 

\section{Discussion and conclusions}\label{disc}
\begin{figure}
\center
\resizebox{\hsize}{!}{\includegraphics[angle=0]{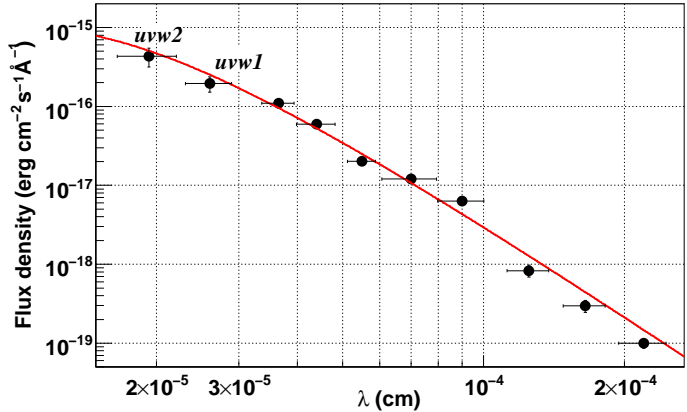}}
\caption{\label{planck} Blackbody fit to the infrared-to-ultraviolet spectral energy distribution of \src. The new \swift\ UV points are labelled with the filter designation.}
\end{figure}
The optical-to-IR spectral energy distribution (SED) of \src\ can be described by a simple Rayleigh--Jeans approximation. With the \swift\ UV points, the blackbody signature and peak location become clearer (see Fig.\,\ref{planck}), allowing us to derive reasonably constrained normalization and temperature [$T \simeq (27\pm8) \times10^3$~K].

A couple of interesting implications follow. The first is that the UV-to-IR SED is well fit by a single blackbody, which suggests that one single component dominates the continuum emission, or, in other words, that the bulk of the UV continuum comes from the same thermal component seen in the optical. It is straightforward to associate this component with the hot surface of the primary WD at a temperature of  $\sim$27\,000 K. The second is that, given this interpretation and the mass range of the primary determined independently in previous studies, the source distance obtained from the blackbody normalization for a primary with radius\footnote{The adopted radius of $\sim$6500 km assumes a primary mass $\sim$0.8 $M_{\odot}$ \citep{nauenberg72}, which is close to the minimum required in the direct impact model (see figure~8 in \citealt{barros07}). The system mass ratio was recently constrained by \citet{roelofs10} to $q = 0.50\pm0.13$, allowing for a lighter primary, $M_1 \simeq 0.55$ $M_{\odot}$ if the orbital evolution were driven only by GW emission. However, the mechanism responsible for the X-ray emission might well affect the orbital evolution, and the allowed range of $q$ does not give a stringent constraint on the primary mass. This latter estimate should thus be taken with caution.\label{massrange}} of $\sim$6500~km is  $D \simeq 0.9\pm0.2~(R_8/6.5)$~kpc (where $R_8$ is the primary radius in units of $10^8$ cm). At this distance, the primary WD can be an important source of irradiation for the secondary with a bolometric (mostly optical) luminosity of $\approx$$1.5 \times 10^{32}(R_8/6.5)^2$~\lum, where half of is directed towards the companion. This is comparable to the X-ray luminosity of the hot spot, $L_{\rm X,\,bol} \approx 5\times 10^{32} (R_8/6.5)^2$~\lum\ (see Sect.\,\ref{xrtspec}).

Besides the intrinsic shape differences, the $uvw1$ and optical modulations are well in phase, suggesting a common origin that we identify with the emission from the surface of the secondary, which is heated by irradiation from the hot primary. The $uvw1$ and optical peaks precede the X-ray modulation by 0.2 or 0.3 in phase if adopting, respectively, the emission peak shifts or the results of the cross-correlation algorithm (see Fig.~\ref{efold}). Therefore, the X-ray-emitting hot spot on the primary surface, responsible for the `saw-tooth' profile of the X-ray pulsations, leads by 0.2/0.3 cycles the line joining the centers of the two stars. 

Whether or not the X-ray emitting hot spot on the primary is also irradiating the secondary depends on its location on the primary surface (and on possible beaming effects). According to the cross-correlation, the angle between the direction connecting the two WDs and the hot spot is $80\degr\pm8\degr$, or possibly $100\degr \pm 10\degr$ if only the offset between the emission peaks is considered. Although the two values are formally compatible, their implications are very different: the smaller value would allow slightly more than half of the secondary face to be irradiated, while the larger value would prevent X-ray irradiation. The result obtained through the cross-correlation essentially agrees with the findings of \citet{roelofs10}, where an angle of about $70\degr$ for the hot spot was suggested. This is based on the location of an He\,\textsc{ii} line emitting bright spot in the Doppler tomogram. On the other hand, the phase shift between the emission peaks in the two bands agrees with the conclusions of \citet{barros07}, who found that the hot spot should lead the line of the centres by more than $90 \degr$ in the direction of the orbital motion. Being based on a more robust measure of the phase-shift, we tend to prefer the former case (no X-ray irradiation from the hot spot) but more observational data are required to better assess this  issue.

While direct impact models naturally predict a phase offset between the impact (X-ray-emitting) spot and the line joining the centers, the observation that optical and X-ray emission have nearly the same offset in the \src\ and RX\,J1914.4+2456, despite the significantly different orbital periods,  orbital separations, and WD masses, requires some degree of fine-tuning. This is well seen from Figure~8 in \citet{barros07}, where the allowed range in parameter space for both systems is shown to be very narrow.

Another challenge for accretion-based models comes from the X-ray luminosity, which we estimated at $\sim$$5\times 10^{32} (R_8/6.5)^2$~\lum\ (see below). The X-ray luminosity could be shifted upwards to $L_{\rm X,\,bol}\simeq10^{33}$~\lum\ (hence, $D\simeq(1.3 \pm 0.3)$~kpc) if we assume the lightest possible primary (0.55~$M_\sun$, cf.~Note\,\ref{massrange}). We conclude that the source luminosity can hardly exceed $10^{33}$~\lum. The models with mass transfer that are able to explain the orbital shrinkage of \src\ (and of RX\,J1914.4+2456) are developed by \citet{dantona06}, \citet{deloye06} and, more recently, \citet{kaplan12}. They all require for \src\ a luminosity greater than $10^{33}$~\lum\ and more likely of the order of $10^{34}$~\lum. For example, orbital shrinkage in the model by \citet{deloye06} is explained if both systems are in a transient phase during which Roche-lobe overflow has `just' started hence, the growing mass transfer rate has not yet reached the value required to balance GW emission. For this phase to be sufficiently long-lived, though, the model requires a relatively large accretion rate, which translates to X-ray luminosities above $10^{33}$--$10^{34}$~\lum\ for both systems (cf. \citealt{deloye06,deloye07}; see also Figs.~4 and 5 in \citealt{kaplan12}, and $\S$3 in \citealt{dantona06}). 

In these accretion-based scenarios, some fine tuning of the stellar parameters would also be required for RX\,J1914.4+2456 to have a larger mass transfer, hence luminosity, given its significantly longer orbital period compared to \src. \citet{dantona06} stressed that their model has difficulties in explaining the properties of RX\,J1914.4+2456 (end of their $\S$3) and postponed further refinements to future investigations. Finally, it is surprising within accretion scenarios that no systems are detected with an orbital period between 5 and 10 minutes but with widening orbits. These would be the descendents of systems like \src. By having a larger luminosity and being in a longer-lived phase of mass tansfer than \src, they should be overabundant. (Considering the small sample of objects like \src, it is, however, difficult to estimate how many such systems should exist and be observable in the Galaxy.)

A phase offset would also be expected in the UI model, which is contrary to what is often asserted, due to the bending of the flux tube caused by the massive flow of currents. Indeed, even on Jupiter, the Io footprint leads the position of the satellite by $\sim$$13^{\circ}$ \citep{goldreich69}. A maximally bent flux tube (cf. \citealt{lai12}) would reach the primary at approximately 90$^{\circ}$ ahead of the line joining the centers, which might explain the similar phase offset in both systems as determined by the saturation of the electric circuit. 

Independent of this, a general efficiency limit of the UI model was recently discussed by \citet{lai12}. For double WD binaries, this translates into a maximum 
luminosity \citep{lai12}:
\begin{equation}
\label{eq:Lai}
L_{\rm max} \approx 10^{31} \zeta_{\phi} \left(\frac{\Delta \Omega}
{\Omega}\right) \mu^2_{32} R^2_9 \left(\frac{{M}_{\mathrm{tot}}}{{M}_{\odot}}\right)^{-5/3} \left(\frac{{P}}{5~{\rm
    min}}\right)^{-13/3}~\mathrm{erg~s}^{-1},
\end{equation}
where $\zeta_{\phi}$ depends on the degree of bending of the flux tube ($\zeta_{\phi}=1$ for a maximally bended tube), $\Delta \Omega$ is the difference between the orbital frequency ($\Omega$) and the primary spin frequency, $\mu$ is the magnetic moment of the primary, $R$ is the secondary radius, $M_{\rm tot}$ the total mass of the system, and $Q_X$ stands for a quantity $Q$ in units of $10^X$. This represents the major problem of the UI model in its current form, since the mechanism cannot provide the required power for the X-ray luminosity of both sources and  the mismatch is significant (although smaller than in the conclusions of \citealt{lai12}, where an extreme value was assumed for the X-ray luminosity of RX\,J1914.4+2456). However, the idealized picture of a DC circuit in the UI model likely needs substantial revision. For instance, if the space between the WDs were not filled with sufficient particles, charge carriers would be stripped from the WD surface and accelerated along field lines. In this case, a radically different physical description would be required (cf. \citealt{alfven58,alfven86}) and the limit of eq.\,(\ref{eq:Lai}) might not apply. Modelling of this scenario is clearly beyond our scope here, but this is a promising direction into which the model should be extended. 

\begin{acknowledgements}
This research is based on observations with the NASA/UKSA/ASI mission \swift. We also used ESO/VLT and the INAF/TNG data. SD acknowledges support from the SFB/Transregio~7, funded by the Deutsche Forschungsgemeinschaft (DFG). We thank the anonymous referee for constructive comments.
\end{acknowledgements}

\bibliographystyle{aa} 
\bibliography{biblio} 

\begin{thebibliography}{41}
\expandafter\ifx\csname natexlab\endcsname\relax\def\natexlab#1{#1}\fi

\bibitem[{{Alfv{\'e}n}(1958)}]{alfven58}
{Alfv{\'e}n}, H. 1958, Tellus, 10, 104

\bibitem[{{Alfv{\'e}n}(1986)}]{alfven86}
{Alfv{\'e}n}, H. 1986, IEEE Transactions on Plasma Science, 14, 779

\bibitem[{{Barros} {et~al.}(2007){Barros}, {Marsh}, {Dhillon}, {Groot},
  {Littlefair}, {Nelemans}, {Roelofs}, {Steeghs}, \& {Wheatley}}]{barros07}
{Barros}, S.~C.~C., {Marsh}, T.~R., {Dhillon}, V.~S., {et~al.} 2007, \mnras,
  374, 1334

\bibitem[{{Breeveld} {et~al.}(2010){Breeveld}, {Curran}, {Hoversten}, {Koch},
  {Landsman}, {Marshall}, {Page}, {Poole}, {Roming}, {Smith}, {Still},
  {Yershov}, {Blustin}, {Brown}, {Gronwall}, {Holland}, {Kuin}, {McGowan},
  {Rosen}, {Boyd}, {Broos}, {Carter}, {Chester}, {Hancock}, {Huckle}, {Immler},
  {Ivanushkina}, {Kennedy}, {Mason}, {Morgan}, {Oates}, {de Pasquale},
  {Schady}, {Siegel}, \& {vanden Berk}}]{breeveld10}
{Breeveld}, A.~A., {Curran}, P.~A., {Hoversten}, E.~A., {et~al.} 2010, \mnras,
  406, 1687

\bibitem[{{Breeveld} {et~al.}(2011){Breeveld}, {Landsman}, {Holland}, {Roming},
  {Kuin}, \& {Page}}]{breeveld11}
{Breeveld}, A.~A., {Landsman}, W., {Holland}, S.~T., {et~al.} 2011, in AIP
  Conference Proceedings, Vol. 1358, Gamma Ray Bursts 2010., ed.
  {J.~E.~McEnery, J.~L.~Racusin, \& N.~Gehrels} (AIP, Melville), 373--376

\bibitem[{{Burrows} {et~al.}(2005){Burrows}, {Hill}, {Nousek}, {Kennea},
  {Wells}, {Osborne}, {Abbey}, {Beardmore}, {Mukerjee}, {Short}, {Chincarini},
  {Campana}, {Citterio}, {Moretti}, {Pagani}, {Tagliaferri}, {Giommi},
  {Capalbi}, {Tamburelli}, {Angelini}, {Cusumano}, {Br{\"a}uninger}, {Burkert},
  \& {Hartner}}]{burrows05}
{Burrows}, D.~N., {Hill}, J.~E., {Nousek}, J.~A., {et~al.} 2005, Space Science
  Reviews, 120, 165

\bibitem[{{Cropper} {et~al.}(1998){Cropper}, {Harrop-Allin}, {Mason}, {Mittaz},
  {Potter}, \& {Ramsay}}]{cropper98}
{Cropper}, M., {Harrop-Allin}, M.~K., {Mason}, K.~O., {et~al.} 1998, \mnras,
  293, L57

\bibitem[{{Cropper} {et~al.}(2004){Cropper}, {Ramsay}, {Wu}, \&
  {Hakala}}]{cropper04}
{Cropper}, M., {Ramsay}, G., {Wu}, K., \& {Hakala}, P. 2004, in Astronomical
  Society of the Pacific Conference Series, Vol. 315, IAU Colloq. 190: Magnetic
  Cataclysmic Variables, ed. {S.~Vrielmann \& M.~Cropper} (ASP, San Francisco),
  324

\bibitem[{{Dall'Osso} {et~al.}(2006){Dall'Osso}, {Israel}, \&
  {Stella}}]{dallosso06}
{Dall'Osso}, S., {Israel}, G.~L., \& {Stella}, L. 2006, \aap, 447, 785

\bibitem[{{Dall'Osso} {et~al.}(2007){Dall'Osso}, {Israel}, \&
  {Stella}}]{dallosso07}
{Dall'Osso}, S., {Israel}, G.~L., \& {Stella}, L. 2007, \aap, 464, 417

\bibitem[{{D'Antona} {et~al.}(2006){D'Antona}, {Ventura}, {Burderi}, \&
  {Teodorescu}}]{dantona06}
{D'Antona}, F., {Ventura}, P., {Burderi}, L., \& {Teodorescu}, A. 2006, \apj,
  653, 1429

\bibitem[{{Deloye} \& {Taam}(2006)}]{deloye06}
{Deloye}, C.~J. \& {Taam}, R.~E. 2006, \apjl, 649, L99

\bibitem[{{Deloye} {et~al.}(2007){Deloye}, {Taam}, {Winisdoerffer}, \&
  {Chabrier}}]{deloye07}
{Deloye}, C.~J., {Taam}, R.~E., {Winisdoerffer}, C., \& {Chabrier}, G. 2007,
  \mnras, 381, 525

\bibitem[{{Fitzpatrick} \& {Massa}(2007)}]{fitzpatrick07}
{Fitzpatrick}, E.~L. \& {Massa}, D. 2007, \apj, 663, 320

\bibitem[{{Goldreich} \& {Lynden-Bell}(1969)}]{goldreich69}
{Goldreich}, P. \& {Lynden-Bell}, D. 1969, \apj, 156, 59

\bibitem[{{Hakala} {et~al.}(2004){Hakala}, {Ramsay}, \& {Byckling}}]{hakala04}
{Hakala}, P., {Ramsay}, G., \& {Byckling}, K. 2004, \mnras, 353, 453

\bibitem[{{Israel} {et~al.}(2004){Israel}, {Covino}, {Dall'Osso}, {Fugazza},
  {Mouche}, {Stella}, {Campana}, {Mangano}, {Marconi}, {Bagnulo}, \&
  {Munari}}]{icd04}
{Israel}, G.~L., {Covino}, S., {Dall'Osso}, S., {et~al.} 2004, Mem. S. A. It.
  Suppl., 5, 148

\bibitem[{{Israel} {et~al.}(2003){Israel}, {Covino}, {Stella}, {Mauche},
  {Campana}, {Marconi}, {Hummel}, {Mereghetti}, {Munari}, \&
  {Negueruela}}]{israel03}
{Israel}, G.~L., {Covino}, S., {Stella}, L., {et~al.} 2003, \apj, 598, 492

\bibitem[{{Israel} \& {Dall'Osso}(2009)}]{israel09}
{Israel}, G.~L. \& {Dall'Osso}, S. 2009, in Astrophysics and Space Science
  Library., Vol. 359, Physics of Relativistic Objects in Compact Binaries: From
  Birth to Coalescence, ed. {M.~Colpi, P.~Casella, V.~Gorini, U.~Moschella, \&
  A.~Possenti } (Springer, Dordrecht), 281

\bibitem[{{Israel} {et~al.}(2005){Israel}, {dall'Osso}, {Mangano}, {Stella},
  {Covino}, {Fugazza}, {Campana}, {Marconi}, {Mereghetti}, \& {Munari}}]{idm05}
{Israel}, G.~L., {dall'Osso}, S., {Mangano}, V., {et~al.} 2005, in AIP
  Conference Series, Vol. 797, Interacting Binaries: Accretion, Evolution, and
  Outcomes, ed. L.~{Burderi}, L.~A. {Antonelli}, F.~{D'Antona}, T.~{di Salvo},
  G.~L. {Israel}, L.~{Piersanti}, A.~{Tornamb{\`e}}, \& O.~{Straniero} (AIP,
  Melville), 307--312

\bibitem[{{Israel} {et~al.}(2002){Israel}, {Hummel}, {Covino}, {Campana},
  {Appenzeller}, {G{\"a}ssler}, {Mantel}, {Marconi}, {Mauche}, {Munari},
  {Negueruela}, {Nicklas}, {Rupprecht}, {Smart}, {Stahl}, \&
  {Stella}}]{israel02}
{Israel}, G.~L., {Hummel}, W., {Covino}, S., {et~al.} 2002, \aap, 386, L13

\bibitem[{{Israel} {et~al.}(1999){Israel}, {Panzera}, {Campana}, {Lazzati},
  {Covino}, {Tagliaferri}, \& {Stella}}]{ipc99}
{Israel}, G.~L., {Panzera}, M.~R., {Campana}, S., {et~al.} 1999, \aap, 349, L1

\bibitem[{{Kaplan} {et~al.}(2012){Kaplan}, {Bildsten}, \&
  {Steinfadt}}]{kaplan12}
{Kaplan}, D.~L., {Bildsten}, L., \& {Steinfadt}, J.~D.~R. 2012, \apj, 758, 64

\bibitem[{{Lai}(2012)}]{lai12}
{Lai}, D. 2012, \apjl, 757, L3

\bibitem[{{Marsh} \& {Steeghs}(2002)}]{marsh02}
{Marsh}, T.~R. \& {Steeghs}, D. 2002, \mnras, 331, L7

\bibitem[{{Mason} {et~al.}(2010){Mason}, {Israel}, {Dall'Osso}, {Stella},
  {Munari}, {Marconi}, {O'Brian}, {Covino}, \& {Fugazza}}]{mason10}
{Mason}, E., {Israel}, G.~L., {Dall'Osso}, S., {et~al.} 2010, eprint:
  astro-ph/1003.1986

\bibitem[{{Nauenberg}(1972)}]{nauenberg72}
{Nauenberg}, M. 1972, \apj, 175, 417

\bibitem[{{Nelemans}(2013)}]{nelemans13}
{Nelemans}, G. 2013, in Astronomical Society of the Pacific Conference Series,
  Vol. 467, The 9th LISA Symposium, ed. G.~{Auger}, P.~{Bin{\'e}truy}, \&
  E.~{Plagnol} (ASP, San Francisco), 27

\bibitem[{{Nelemans} {et~al.}(2001){Nelemans}, {Portegies Zwart}, {Verbunt}, \&
  {Yungelson}}]{nelemans01}
{Nelemans}, G., {Portegies Zwart}, S.~F., {Verbunt}, F., \& {Yungelson}, L.~R.
  2001, \aap, 368, 939

\bibitem[{{Poole} {et~al.}(2008){Poole}, {Breeveld}, {Page}, {Landsman},
  {Holland}, {Roming}, {Kuin}, {Brown}, {Gronwall}, {Hunsberger}, {Koch},
  {Mason}, {Schady}, {vanden Berk}, {Blustin}, {Boyd}, {Broos}, {Carter},
  {Chester}, {Cucchiara}, {Hancock}, {Huckle}, {Immler}, {Ivanushkina},
  {Kennedy}, {Marshall}, {Morgan}, {Pandey}, {de Pasquale}, {Smith}, \&
  {Still}}]{poole08}
{Poole}, T.~S., {Breeveld}, A.~A., {Page}, M.~J., {et~al.} 2008, \mnras, 383,
  627

\bibitem[{{Predehl} \& {Schmitt}(1995)}]{predehl95}
{Predehl}, P. \& {Schmitt}, J.~H.~M.~M. 1995, \aap, 293, 889

\bibitem[{{Ramsay} {et~al.}(2002){Ramsay}, {Hakala}, \& {Cropper}}]{ramsay02}
{Ramsay}, G., {Hakala}, P., \& {Cropper}, M. 2002, \mnras, 332, L7

\bibitem[{{Roelofs} {et~al.}(2010){Roelofs}, {Rau}, {Marsh}, {Steeghs},
  {Groot}, \& {Nelemans}}]{roelofs10}
{Roelofs}, G.~H.~A., {Rau}, A., {Marsh}, T.~R., {et~al.} 2010, \apjl, 711, L138

\bibitem[{{Roming} {et~al.}(2005){Roming}, {Kennedy}, {Mason}, {Nousek}, {Ahr},
  {Bingham}, {Broos}, {Carter}, {Hancock}, {Huckle}, {Hunsberger}, {Kawakami},
  {Killough}, {Koch}, {McLelland}, {Smith}, {Smith}, {Soto}, {Boyd},
  {Breeveld}, {Holland}, {Ivanushkina}, {Pryzby}, {Still}, \&
  {Stock}}]{roming05}
{Roming}, P.~W.~A., {Kennedy}, T.~E., {Mason}, K.~O., {et~al.} 2005, Space
  Science Reviews, 120, 95

\bibitem[{{Schlegel} {et~al.}(1998){Schlegel}, {Finkbeiner}, \&
  {Davis}}]{schlegel98}
{Schlegel}, D.~J., {Finkbeiner}, D.~P., \& {Davis}, M. 1998, \apj, 500, 525

\bibitem[{{Solheim}(2010)}]{solheim10}
{Solheim}, J.-E. 2010, \pasp, 122, 1133

\bibitem[{{Strohmayer}(2003)}]{strohmayer03}
{Strohmayer}, T.~E. 2003, \apjl, 593, L39

\bibitem[{{Strohmayer}(2005)}]{strohmayer05}
{Strohmayer}, T.~E. 2005, \apj, 627, 920

\bibitem[{{Strohmayer}(2008)}]{strohmayer08}
{Strohmayer}, T.~E. 2008, \apjl, 679, L109

\bibitem[{{Wu}(2009)}]{wu09}
{Wu}, K. 2009, Research in Astronomy and Astrophysics, 9, 725

\bibitem[{{Wu} {et~al.}(2002){Wu}, {Cropper}, {Ramsay}, \& {Sekiguchi}}]{wu02}
{Wu}, K., {Cropper}, M., {Ramsay}, G., \& {Sekiguchi}, K. 2002, \mnras, 331,
  221

\end{thebibliography}

\end{document}